\documentclass[a4paper,11pt]{article}
\pdfoutput=1
\usepackage[utf8]{inputenc}
\usepackage{amssymb}
\usepackage{amsmath}
\usepackage{amsthm}
\usepackage{amsfonts}
\usepackage[pdftex]{graphicx}
\usepackage{geometry}
\usepackage{braket}
\usepackage{enumerate}
\usepackage{comment}
\usepackage{lscape}
\usepackage{graphicx}
\usepackage{amssymb}
\usepackage{amsmath}
\usepackage{bbm}
\usepackage{cancel}
\usepackage{epstopdf}
\usepackage{mathtools}
\usepackage{braket}
\usepackage[export]{adjustbox}
\usepackage[usenames,dvipsnames]{xcolor}
\usepackage{tikz}
\usetikzlibrary{cd}
\usepackage{relsize}
\usepackage{titlesec}
\usepackage{setspace} 
\usepackage{xpatch}
\usepackage{lipsum} 
\usepackage{xfrac}
\usepackage[final]{hyperref}
\hypersetup{
    colorlinks=true,
    linkcolor=blue,
    urlcolor=blue,
    linktoc=all
}

\usepackage{setspace}
\usepackage[in]{fullpage}
\usepackage{hyperref}
\usepackage{array}
\newcolumntype{P}[1]{>{\centering\arraybackslash}p{#1}}
\newcolumntype{M}[1]{>{\centering\arraybackslash}m{#1}}
\usepackage{cancel}
\usepackage{wrapfig}
\usepackage[font=small,labelfont=bf]{caption}
\usepackage{pifont}

\usepackage[dvipsnames]{xcolor}
\usepackage{mathtools}

\usepackage{color}

\usepackage{tikz}
\usetikzlibrary{arrows,shapes}
\usetikzlibrary{trees}
\usetikzlibrary{matrix,arrows} 
\usetikzlibrary{positioning}
\usetikzlibrary{calc,through}
\usetikzlibrary{decorations.pathreplacing}
\usepackage{pgffor}	

\numberwithin{equation}{section}

\usepackage[english]{babel}

\usepackage{caption}

\usepackage[nottoc,notlot,notlof]{tocbibind}
\usepackage[nosort]{cite}   
\usepackage{color}

\usepackage{parskip}
\usepackage[T1]{fontenc}

\usepackage{lmodern}
\usepackage{yfonts}
\usepackage{bbm}
\usepackage{physics}
\usepackage{slashed}
\usepackage{makeidx}
\usepackage[vcentermath]{youngtab}
\usepackage{amsthm}
\usepackage[scr=boondoxo]{mathalfa}

\hypersetup{
	colorlinks=true,
	linktoc=page,
	citecolor=blue,
	linkcolor=blue,
	urlcolor=blue} 

\makeatletter \def\@captype{figure} \makeatother 
\usetikzlibrary{calc}

\makeatletter

\begin{document}

\begin{flushright}
	 QMUL-PH-23-21\\
\end{flushright}

\vspace{20pt} 

\begin{center}

	{\Large \bf  {Self-Dual Fields on Self-Dual Backgrounds\\ and the Double Copy} }  \\
	\vspace{0.3 cm}

	\vspace{20pt}

	{\mbox {\sf  \!\!\!\! Graham~R.~Brown, Joshua~Gowdy, Bill~Spence
	}}
	\vspace{0.5cm}

	\begin{center}
		{\small \em
			Centre for Theoretical Physics\\
			Department of Physics and Astronomy\\
			Queen Mary University of London\\
			Mile End Road, London E1 4NS, United Kingdom
		}
		
	\end{center}

	\vspace{40pt}  

	{\bf Abstract}
\end{center}

\vspace{0.3cm}

\noindent

We explore the double copy for self-dual gauge and gravitational fields on  self-dual background spacetimes. We consider backgrounds associated to solutions of the second Plebanski equation and  describe results with different gauge-fixing conditions. Finally we discuss the kinematic and $w$-algebras and the double copy, identifying modified Poisson structures and kinematic structure constants in the presence of the self-dual background.  The self-dual plane wave and Eguchi-Hanson spacetimes are studied as examples and their respective $w$-algebras derived.
 
\vfill
\hrulefill
\newline
\vspace{-1cm}
\!\!{\tt\footnotesize\{graham.brown, j.k.gowdy, w.j.spence\}@qmul.ac.uk}

\setcounter{page}{0}
\thispagestyle{empty}
\newpage


\setcounter{tocdepth}{4}
\hrule height 0.75pt
\tableofcontents
\vspace{0.8cm}
\hrule height 0.75pt
\vspace{1cm}
\setcounter{tocdepth}{2}

\section{Introduction}
\label{IntroductionSection}

The study of self-dual gauge and gravitational fields has provided a fertile source of ideas and results in physics and mathematics. Work some time ago showed that self-dual theories in a light-cone gauge  could be described by equations for scalar fields \cite{Plebanski:1975wn,
Bardeen:1995gk,Chalmers:1996rq,Prasad:1979zc,Dolan:1983bp,Parkes:1992rz,Cangemi:1996pf,Popov:1996uu,Popov:1998pc} corresponding to the positive helicity sectors of Yang-Mills and gravity. The self-dual sector allows for a simplified study of many features of the full theories. An area of recent interest is the investigation of self-dual fields in order to learn more about the structure of the double copy\footnote{The classical double copy for solutions  of the equations of motion was first explored in \cite{Monteiro:2014cda,Luna:2015paa}} and kinematic algebras (see \cite{Chacon:2020fmr,Campiglia:2021srh,Monteiro:2022nqt,Monteiro:2013rya,Cheung:2016prv,Chen:2019ywi,
 Brandhuber:2021bsf,Brandhuber:2022enp,
  Elor:2020nqe,Armstrong-Williams:2022apo,Farnsworth:2021wvs,Skvortsov:2022unu,He:2015wgf,Berman:2018hwd,Nagy:2022xxs,Farnsworth:2023mff,Easson:2023dbk, Borsten:2021hua,Borsten:2021gyl,Borsten:2022vtg,Borsten:2023ned,Borsten:2023paw} for  recent more general work on the double copy and CK duality). In \cite{Monteiro:2011pc} it was shown that self-dual Yang-Mills and gravity have manifest colour-kinematics duality, and the kinematic algebra was identified as that of area-preserving diffeomorphisms of the plane. In the context of celestial holography\cite{Pasterski:2016qvg,Pasterski:2017kqt,Pasterski:2017ylz,Raclariu:2021zjz,Pasterski:2021rjz,Pasterski:2021raf,McLoughlin:2022ljp} this algebra was shown appear through the soft and collinear limit of positive helicity gravitons as (the wedge subalgebra of) $w_{1+\infty}$ \cite{Strominger:2021lvk}. This link between kinematic algebras and OPEs in celestial holography \cite{Fan:2019emx,Pate:2019lpp,Bhardwaj:2022anh,Guevara:2021tvr,Banerjee:2020kaa,Adamo:2022wjo,Ren:2023trv,Guevara:2021abz,Mago:2021wje,Ren:2022sws,Ball:2022bgg, Banerjee:2020zlg,Ebert:2020nqf,Banerjee:2020vnt} was recently discussed in detail in \cite{Monteiro:2022lwm}.

  A natural generalisation of this is to study self-duality conditions on non-flat spacetimes. Some progress has been made in understanding how the double copy can be applied to curved backgrounds \cite{Adamo:2017nia,Adamo:2018mpq,Bahjat-Abbas:2017htu,Carrillo-Gonzalez:2017iyj,Borsten:2019prq,Borsten:2021zir,Chawla:2022ogv, Costello:2022jpg, Costello:2023hmi}, with the case of AdS receiving particular attention
\cite{Armstrong:2020woi,Albayrak:2020fyp,Alday:2021odx,Diwakar:2021juk,Sivaramakrishnan:2021srm,Cheung:2022pdk,Herderschee:2022ntr,Drummond:2022dxd,Farrow:2018yni,Lipstein:2019mpu,Jain:2021qcl,Zhou:2021gnu,Armstrong:2022csc, Lipstein:2023pih,Armstrong:2023phb,Mei:2023jkb}.

In this letter we would like to study self-duality for  the case of {\it self-dual} spacetime backgrounds (see \cite{Costello:2022jpg, Costello:2023hmi,Adamo:2020yzi,Adamo:2022mev,Bittleston:2023bzp} for some recent work on this topic). We will work with the spacetime metrics defined by solutions of the second Plebanski equation, and study the conditions for the existence of self-dual Yang-Mills fields, and self-dual metric variations, on these backgrounds. 
In the flat space case, dealt with in section \ref{sec: flat}, there are two formulations of the self-duality conditions which are related by a simple relabelling of coordinates. A general self-dual background (Yang-Mills backgrounds are dealt with in section  \ref{sec: generalSDYMBackground} and gravity backgrounds in section \ref{sec: generalSDBackground})  does not have this symmetry, and we find that these two formulations generalise quite differently. 
The first class of solutions in gravity backgrounds, which we call  a \lq matched\rq\ gauge,  can be seen as  generalising the flat space solution to curved self-dual backgrounds by linearly perturbing the Plebanski scalar. We also find a second class of solutions, which we call a \lq flipped\rq\ gauge, which requires a Kerr-Schild condition on the background and leads to a modified Poisson structure coming from the Plebanski equation in this gauge. We then describe aspects of the double copy, and kinematic and $w$-algebras revealed by these results. 

This general formulation is discussed in detail in two examples -  the self-dual plane wave spacetime in section \ref{sec: SDPWBackground}
and the Eguchi-Hanson (EH) metric in section \ref{sec: EHBackground}. These case studies connect with some of the  results developed recently in twistor space in  \cite{Adamo:2022mev} and \cite{Bittleston:2023bzp}. 
For the self-dual plane wave background we find that a natural definition of a \lq plane wave\rq-like solution to the wave/Plebanski equation in that spacetime leads to a kinematic algebra with modified structure constants when compared to the flat background. Nevertheless these structure constants match the flat-space case in the holomorphic collinear limit of the two \lq plane wave\rq\,  solutions and so generate the standard flat-space $w$-algebra. The soft generators generating the algebra are however altered and correspond to the expansion of the particular \lq plane wave\rq\, solutions adapted to the self-dual plane wave background. The double copy in our formulation replaces the Lie algebra commutators with Poisson brackets and leads to a so-called double bracket in the Plebanski equation. When acting on \lq plane wave\rq\, solutions, in flipped gauge, we show this procedure replaces colour structure constants with those from the kinematic algebra $X_{PW}(k_1,k_2)$ and so gives the expected squaring relation of the single copy, i.e. $X_{PW}(k_1,k_2)^2$.

In the Eguchi-Hanson background, we express the more complicated solutions of the wave equation discussed in \cite{Bittleston:2023bzp} in spacetime coordinates. We then show that the Poisson bracket of two of these \lq plane waves\rq\,  gives an expression for the deformed kinematic structure constants $X_{EH}(k_1,k_2)$, which we define in the holomorphic collinear limit. The double bracket of two \lq plane waves\rq\, is then shown to give the square of this expression, but with additional terms, demonstrating that
 in the  Eguchi-Hanson background the kinematic algebra squaring relations are modified by curvature terms. Since even in the holomorphic collinear limit $X_{EH}(k_1,k_2)$ differs from the flat-space and self-dual plane wave cases, we then expect a completely different \lq $w$-algebra\rq\,  of soft generators. We derive this algebra of soft generators following the same method as in the previous cases, by expanding the \lq plane wave\rq\, solutions, giving a spacetime realisation of  the results of \cite{Bittleston:2023bzp} coming from twistor space.


\section{Flat background}\label{sec: flat}
We start by setting notation and briefly recalling  the standard results for self-dual Yang-Mills (YM) and self-dual gravity in a flat background. In this section, we will generally follow the discussion in \cite{Monteiro:2011pc}.
The spacetime coordinates are taken to be $(u,v,X,Y)$, with the metric 
\begin{equation}\label{eq:Flatmetric}
ds^2 = 2 du\, dv - 2 dX dY .
\end{equation}
For real coordinates, this implies we are using $(2,2)$ signature. The coordinates $(u,v,X,Y)$ are related to the usual $(t,x,y,z)$ as follows
\begin{equation}
    \begin{split}
        u=\frac{t+z}{\sqrt{2}},\quad v=\frac{t-z}{\sqrt{2}},\quad X= \frac{x+y}{\sqrt{2}},\quad Y=\frac{x-y}{\sqrt{2}}
    \end{split}
\end{equation} 
and in terms of the coordinates $(t,x,y,z)$ the metric signature is $(+,-,+,-)$.


\subsection{Self-dual Yang-Mills}
A gauge field $A_\mu=(A_u,A_v,A_X,A_Y)$ on flat space with metric \eqref{eq:Flatmetric}
is self-dual if its field strength\footnote{In our conventions $ F_{\mu\nu}= \partial_{\mu}A_{\nu}-\partial_{\nu}A_{\mu}+[A_{\mu},A_{\nu}]$.} satisfies
\begin{equation}
    F_{\mu\nu}= \frac{\sqrt{g}}{2}\epsilon_{\mu\nu\rho\sigma}F^{\rho\sigma}\,.
\end{equation}
where $g$ is the determinant of the metric \eqref{eq:Flatmetric}.
Imposing the gauge-fixing condition $A_v=0$, the self-duality condition above can be satisfied by setting $A_X=0$ and 
\begin{equation}\label{eq:Asolsflat}
\begin{split}
A_u &=  \phi_X ,\\
A_Y &=  \phi_v , 
\end{split}
\end{equation}
for a (Lie algebra valued) function $\phi(u,v,X,Y)$ satisfying the self-dual Yang-Mills equation
\begin{equation}\label{eq:PlebYMflat}
\Box\phi -2[\phi_v,\phi_X] = 0 ,
\end{equation}
where the scalar Laplacian is $\Box = 2(\partial_u\partial_v-\partial_X\partial_Y)$.
In what follows, we will be using a notation where subscripts on  scalar fields such as $\phi$  signify partial derivatives. For example, $\phi_v=\partial_v\phi, \phi_{Xv}=\partial_X\partial_v\phi$ - this should not be confused with the use of subscripts to denote components of covectors, for example $k_\mu=(k_u,k_v,k_X,k_Y)$.
If we introduce the Poisson bracket
\begin{equation}\label{eq:PoissonBYMflat}
\{f,g\} = \partial_v f \partial_X g   -\partial_X f \partial_v g ,
\end{equation}
then the self-dual Yang-Mills equation becomes
\begin{equation}\label{eq:PlebYMflatPB}
\Box\phi -[\{\phi,\phi\}] = 0 ,
\end{equation}
where we have used a notation suggestive of colour-kinematics duality, as used in \cite{Lipstein:2023pih},
\begin{equation}\label{compoisbracket}
    [\{f,g\}]:=[f_v,g_X]-[f_X,g_v] \,.
\end{equation}
For the covector $k_\mu=(k_u,k_v,k_X,k_Y)$ and coordinate vector $x^\mu=(u,v,X,Y)$, with $k\cdot x:=k_\mu x^\mu$, the plane wave $e^{i k\cdot x}$ satisfies
\begin{equation}\label{eq:planewave}
\Box\, e^{i k\cdot x} = 0 
\end{equation}
if $k_\mu$ is a null vector. 
In momentum space, the cubic coupling arising from the self-dual YM equation \eqref{eq:PlebYMflatPB} involves the kinematic structure constants
\begin{equation}\label{eq:cubicX}
X(k_1,k_2) = k_{1X} k_{2v}-k_{1v}k_{2X} ,
\end{equation}
along with the Lie algebra constants $f^{abc}$.
Explicitly, the bracket \eqref{compoisbracket} of two plane waves and Lie algebra generators satisfies\cite{Monteiro:2011pc}
\begin{equation}\label{eq:cubicX2}
[\{T^a e^{i k_1\cdot x} , T^b e^{i k_2\cdot x}  \}] =  X(k_1,k_2)f^{abc}\, T^{c} e^{i (k_1+k_2)\cdot x} .
\end{equation}

There is also an alternative  gauge-fixing condition $A_u=0$, for which the self-duality condition  can be satisfied by setting $A_Y=0$ and 
\begin{equation}\label{eq:Asolsflat2}
\begin{split}
A_v &=  \phi_Y ,\\
A_X &=  \phi_u , 
\end{split}
\end{equation}
for a function $\phi(u,v,X,Y)$ satisfying the self-dual YM equation \eqref{eq:PlebYMflat} but with the coordinates $u\leftrightarrow v$ and $Y\leftrightarrow X$ exchanged, that is
\begin{equation}\label{eq:PlebYMflatnewgauge}
\Box\phi -2[\phi_u,\phi_Y] = 0 .
\end{equation}
The flat metric is invariant under this exchange and so results obtained in this new gauge are  trivially related to the previous gauge by a simple interchange of coordinates. This is not the case when we consider self-dual backgrounds in the sections below, since these backgrounds have no such symmetry, and we will describe the two different gauges separately.


\subsection{Self-dual gravity}
Now we recall the analogous construction for self-dual gravity, where the metric is taken to be the following variation from the flat metric:
\begin{equation}\label{eq:Gravmetric}
ds^2 = g_{\mu\nu}(\Psi)dx^\mu dx^\nu =2 du dv - 2 dX dY + \Psi_{XX} du^2 + \Psi_{vv} dY^2 + 2 \Psi_{Xv} du dY ,
\end{equation}
with some function $\Psi(u,v,X,Y)$. 
Define the expression
\begin{equation}\label{eq:GravPlebForm}
{\rm Pleb}_0(\Psi) :=2( \Psi_{uv}-\Psi_{XY}) - \Psi_{XX}\Psi_{vv} + (\Psi_{vX})^2,
\end{equation}
where the subscript $0$ indicates the flat background, and define the operator $\Delta_0(\Psi)$ by the variation of this expression as 
\begin{equation}\label{eq:GravPleb}
 \Delta_0(\Psi)(\delta\Psi) := \delta({\rm Pleb}_0(\Psi)) .
\end{equation}
Explicitly
\begin{equation}\label{eq:GravOp}
\Delta_0(\Psi) = 2(\partial_{uv}-\partial_{XY} )-\Psi_{XX}\partial_{vv} + 2 \Psi_{vX}\partial_{vX} - \Psi_{vv}\partial_{XX}  .
\end{equation}
Then the anti-self-dual part of the Weyl tensor is zero except for the component $C^-_{uYuY}$ (and components related to this by the symmetries of the tensor), and we find
\begin{equation}\label{eq:GravASDWeyl}
C^-_{uYuY}=  -\frac{1}{4}\Delta_0(\Psi) {\rm Pleb}_0(\Psi) .
\end{equation}
The non-vanishing components of the Ricci tensor are given by
\begin{equation}\label{eq:RicciFlat}
R_{ab}=  -\frac{1}{2}\, \partial_{\bar a}\partial_{\bar b}{\rm Pleb}_0(\Psi) .
\end{equation}
where $a,b=(u,Y)$ and $\bar u= X, \bar Y = v$. Thus the metric $g_{\mu\nu}(\Psi)$ given by \eqref{eq:Gravmetric} is Ricci-flat and has self-dual Weyl tensor if the scalar field $\Psi$ satisfies the gravitational Plebanski equation 
\begin{equation}\label{eq:GravPlebSDG}
{\rm Pleb}_0(\Psi)= 2(\Psi_{uv}-\Psi_{XY}) -  \Psi_{XX}\Psi_{vv} + (\Psi_{vX})^2 = 0 .
\end{equation}
Defining the following gravitational  bracket $\{\{ \cdot ,\cdot\}\}$ using \eqref{eq:PoissonBYMflat}
\begin{equation}\label{eq:PoissonBgrav}
\{\{f,g\}\} = \frac{1}{2}\Big(\{\partial_v f, \partial_X g\}   -\{\partial_X f,  \partial_v g\}\Big) \,,
\end{equation}
the Plebanski equation \eqref{eq:GravPlebSDG} can be written
\begin{equation}\label{eq:PlebGravPB}
\Box\Psi -\{\{\Psi,\Psi\}\} = 0 ,
\end{equation}
revealing the double copy relation \cite{Monteiro:2011pc} $\phi\rightarrow\Psi,$ $[\{\cdot,\cdot\}]\rightarrow\{\{\cdot,\cdot\}\}$ compared with \eqref{eq:PlebYMflatPB}. Furthermore, we can consider the gravitational bracket acting on a pair of plane wave solutions in flat space and we find the following  double copy structure
\begin{equation}\label{eq:Gravbracketeikx}
\{\{e^{i k_1\cdot x}, e^{i k_2\cdot x} \}\} = \frac{1}{2} e^{i (k_1+k_2)\cdot x} X(k_1,k_2)^2\, ,
\end{equation}
where the colour structure constants in \eqref{eq:cubicX2} have been replaced by additional kinematic ones. Alternatively, from \eqref{eq:cubicX2} one may strip off the colour structure and isolate the kinematic algebra as the Poisson bracket of two plane waves
\begin{equation}\label{eq: kinematicAlgebra}
    \{e^{i k_1\cdot x}, e^{i k_2\cdot x} \} =  e^{i (k_1+k_2)\cdot x} X(k_1,k_2)\, .
\end{equation}

As explained in \cite{Monteiro:2022lwm}, in the context of celestial holography the appearance of the \lq left structure constants\rq\ $X(k_1,k_2)$ in both YM and gravity implies the chirality of the operator product expansion in both cases. The second \lq right structure constants\rq\ , $f^{abc}$ in the YM case and the second copy of  
$X(k_1,k_2)$ in the gravity case, correspond to the structure constants of the OPEs. The soft expansion of the latter may be explored by noting that
the null vector condition $k^2=2(k_uk_v-k_Xk_Y)=0$ implies that
 we may set 
\begin{equation}\label{eq:ratios}
\frac{ k_u}{k_X} = \frac{k_Y}{k_v} = \rho
\end{equation}
for some $\rho$. Thus we can write
\begin{equation}\label{eq:kdotx}
k \cdot x = (\rho Y+ v) k_v + (\rho u + X ) k_X \,.
\end{equation}
The soft limit of the momentum $k$ then corresponds to $(k_v,k_x)\rightarrow0$ at fixed $\rho$. Expanding 
 $e^{i k\cdot x}$ in this limit gives
\begin{equation}\label{eq:eikx}
e^{i k\cdot x} = \sum_{a,b=0}^{\infty} \frac{(ik_v)^a(ik_X)^b}{a!b!} e_{ab} ,
\end{equation}
where the \lq soft mode generators\rq\ are given by $e_{ab}= (\rho Y+v) ^a (\rho u + X)^b$. To make contact with the algebras appearing in celestial holography, we now need to take the collinear limit of the two momentum $k_1,$ $k_2$ appearing in the algebra \eqref{eq: kinematicAlgebra}. To do this we use the holomorphic collinear limit where $(\rho_1-\rho_2)\rightarrow 0$. This makes $k_1$ and $k_2$ collinear since 
\begin{equation}
   k_1\cdot k_2= (\rho_1-\rho_2)X(k_1,k_2)\, .
\end{equation}
At leading order in the holomorphic collinear limit (corresponding to the first term in the OPE expansion in the celestial holography context)  we may set $\rho_1=\rho_2=\rho$ and substitute the expansion \eqref{eq:eikx} into the kinematic algebra \eqref{eq: kinematicAlgebra} to obtain
\begin{equation}\label{eq:egenPB}
\{e_{a,b},e_{c,d}\} = ( ad-bc) e_{a+c-1,b+d-1} .
\end{equation}

Defining the conventional generators
$w^p_m=\frac{1}{2}e_{p-1+m,p-1-m}$ we then find the wedge sub-algebra of the $w_{1+\infty}$ algebra
\begin{equation}\label{eq:wPB}
\{w_m^p,w_n^q\} = \Big( m(q-1)-n(p-1)\Big) w_{m+n}^{p+q-2} .
\end{equation}
The conditions that $a$ and $b$ are integers greater than or equal to zero translates to the conditions that $p,m$ are half-integers and satisfy $1-p\leq m\leq p-1$ and $p\geq1$ (similarly for $q,n$). This algebra has been studied  in the celestial holography context
\cite{Guevara:2021abz, Strominger:2021lvk,Ball:2021tmb} where it is generated by the commutation relations of operators inserting soft gravitons.

As in the YM case, there is  another gauge-fixing condition related to the above by interchanging $u\leftrightarrow v$ and $Y\leftrightarrow X$, with (trivially equivalent) consequent equations. We reiterate that these different types of gauge will not be as trivially related once we consider self-dual backgrounds in the next section.

\section{General self-dual YM backgrounds}\label{sec: generalSDYMBackground}

Our first exploration of self-dual perturbations of self-dual backgrounds starts with YM backgrounds in flat space. To begin with we consider a background self-dual gauge field $A(\chi)$ in the gauge \eqref{eq:Asolsflat}
\begin{equation}\label{eq: flatYMbackground}
    A(\chi)=(\chi_X,0,0,\chi_v)\,,\quad \Box\chi -[\{\chi,\chi\}] = 0 \,.
\end{equation}
Since the gauge field is linear in the scalar $\chi$ we can write a self-dual perturbation on this background as
\begin{equation}\label{eq: YMbackYMpertubation}
    A(\chi+\psi)=(\chi_X+\psi_X,0,0,\chi_v+\psi_v)= A(\chi)+ A(\psi)\,.
\end{equation}
where $A(\psi)$ is the perturbation. The total gauge field $A(\chi+\psi)$  must then satisfy the self-dual YM equation \eqref{eq:PlebYMflat}
\begin{align}\label{eq: YMplebanskiYMBackground}
    &\Box(\chi+\psi) -[\{\chi+\psi,\chi+\psi\}] \nonumber\\
    &=\widetilde{\Box}_{\chi}(\psi) -[\{\psi,\psi\}] =0\,,
\end{align}
where we have used \eqref{eq: flatYMbackground} and defined a ``deformed'' scalar Laplacian 
\begin{equation}\label{eq: deformedLaplacian}
    \widetilde{\Box}_{\chi}= \Box -2[\{\chi,\cdot\}]=D_\chi^\mu D_{\chi\,\mu}\,,
\end{equation}
which is simply the scalar Laplacian in the background gauge field $A(\chi)$. The covariant derivative is given by\footnote{This is consistent with our conventions for the field strength since $[D_\mu,D_\nu]\psi=[F_{\mu\nu},\psi]$ for adjoint valued fields $\psi$.}\begin{equation}
    D_{\chi}^\mu\coloneqq \partial^{\mu}+[A^\mu(\chi),\cdot]\,,
\end{equation} 
when acting on an adjoint valued field.
Equation \eqref{eq: YMplebanskiYMBackground} is the analogue of  \eqref{eq:PlebYMflat} in a  self-dual background YM field. 

We will see shortly that the discussion above can be double copied in two ways. First, we can just double copy the background gauge field to obtain equations of motion for self-dual YM on a self-dual gravitational background (Sec. \ref{sec: YMPertubationGRbackground}). Second, we can double copy  both the background and the perturbation, to obtain self-dual gravity perturbations on a self-dual background (Sec. \ref{sec: gravityPertubation}). In either case, the scalar Laplacian \eqref{eq: deformedLaplacian} will double copy to a familiar object.
%
\section{General self-dual background spacetimes}\label{sec: generalSDBackground}
We now turn to generalising the above results in section \ref{sec: flat} valid for flat backgrounds to the case of self-dual background metrics. This leads us to the two possible double copies of the case considered in section \ref{sec: generalSDYMBackground} of self-dual YM fields on self-dual YM backgrounds, these are summarised in the diagram in \eqref{fig: double copies}.

We consider self-dual metrics of the form 
\begin{equation}\label{eq:GenMetric}
ds^2= g_{\mu\nu}(\Phi) dx^\mu dx^\nu = 2 du dv - 2 dX dY + \Phi_{XX} du^2 + \Phi_{vv} \,dY^2 + 2 \Phi_{Xv} \,du\, dY
\end{equation}
for a scalar function $\Phi(u,v,X,Y)$ satisfying the Plebanksi equation ${\rm Pleb}_0(\Phi)=0$.

Given a co-vector $k_\mu=(k_u,k_v,k_X,k_Y)$ we have
\begin{equation}\label{eq:ksq}
\begin{split}
k^\mu k_\mu &= 2k_u k_v - 2 k_X k_Y - k_v^2 \Phi_{XX} -k_X^2 \Phi_{vv} + 2 k_v k_X \Phi_{Xv} \\
& = 2 \hat k_u \hat k_v -2 \hat k_X \hat k_Y\,,
\end{split}
\end{equation}
where it proves useful to define
\begin{equation}\label{eq:shiftks}
\begin{split}
&\hat k_u := k_u -\frac{1}{2}\Big(k_v \Phi_{XX} - k_X\Phi_{Xv}\Big) ,\quad \hat k_v := k_v , \\
&\hat k_X := k_X , \quad \hat k_Y:= k_Y -\frac{1}{2}\Big(k_v \Phi_{Xv} - k_X\Phi_{vv}\Big)\, .
\end{split}
\end{equation}
The hatted momenta are those in the tangent space - if we write the vierbein
\begin{equation}\label{eq:vierbein}
\begin{split}
e_u^a &= (1,\frac{1}{2}\Phi_{XX},-\frac{1}{2}\Phi_{Xv},0),  \quad
e_v^a = (0,1,0,0),  \\
e_X^a &= (0,0,1,0),  \quad
e_Y^a = (0,\frac{1}{2}\Phi_{Xv},-\frac{1}{2}\Phi_{vv},1),
\end{split}
\end{equation}
satisfying $e_\mu^a e_{\nu a} = g_{\mu\nu}(\Phi)$, with $g_{\mu\nu}(\Phi)$ the metric in \eqref{eq:GenMetric},  then $\hat k_a= e_a^\mu k_\mu$.

\subsection{Self-dual Yang-Mills}\label{sec: YMPertubationGRbackground}
With a gauge field $A_\mu=(A_u,A_v,A_X,A_Y)$ on this spacetime we can choose the  gauge-fixing condition $n^\mu A_\mu=0$, with null vector  $n^\mu=(0,1,0,0)$, thus setting
\begin{equation}
A_v=0 .
\end{equation}
Now we require that the field strength $F_{\mu\nu}$ is self-dual, i.e. that the anti-self-dual components $F^-_{\mu\nu}$ vanish. This imposes three independent conditions. Two of these are satisfied if we set 
\begin{equation}\label{eq:Asols1}
\begin{split}
A_X & =0 , \\
A_u &=  \phi_X , \\
A_Y &=  \phi_v , 
\end{split}
\end{equation}
for a scalar field $\phi(u,v,X,Y)$, and the final self-duality condition imposes the equation
\begin{equation}\label{eq:GenMetPleb}
\Box_{\Phi}\phi -2[\phi_v,\phi_X] = 0 ,
\end{equation}
where $\Box_{\Phi}$ is the Laplacian in the metric \eqref{eq:GenMetric}. This gauge matches the choice made for the self-dual background metric, wherein the components of the metric $g_{\mu\nu}$ satisfy $g_{vv}=g_{vX}=g_{XX}=0$; we call this gauge the `\textit{matched gauge}'.
Equation \eqref{eq:GenMetPleb} is thus the generalisation of the self-dual Yang-Mills equation \eqref{eq:PlebYMflat} to the  background \eqref{eq:GenMetric}. If we define the Poisson bracket as in the flat space case
\begin{equation}\label{eq:PoissonBgen}
\{f,g\} = \partial_v f \partial_X g  - \partial_v g \partial_X f \, ,
\end{equation}
then the Plebanski equation \eqref{eq:GenMetPleb} can  be written as
\begin{equation}\label{eq:PlebYMEHPB}
\Box_{\Phi}\phi -[\{\phi,\phi\}] = 0 \,,
\end{equation}
with $[\{\phi,\phi\}]$ defined in \eqref{compoisbracket}. The equation above can be viewed as the double copy of \eqref{eq: YMplebanskiYMBackground} where we only double copy the background gauge field. Explicitly, performing the double copy  on only the background $\chi$ using  $[\{\chi,\cdot\}]\rightarrow\{\{\Phi,\cdot\}\}$, the scalar Laplacian in gauge theory $\widetilde{\Box}_{\chi}$ \eqref{eq: deformedLaplacian} becomes
\begin{equation}
    \widetilde{\Box}_{\chi}= \Box -2[\{\chi,\cdot\}]\rightarrow \Box -2\{\{\Phi,\cdot\}\}= \Box_{\Phi}
\end{equation}
where in the last equality we combined $-2\{\{\Phi,\cdot\}\}$ with the flat scalar Laplacian to give us the curved Laplacian on the background $\Phi$.

As we have noted earlier, there is also a different gauge choice which reduces to the  gauge-fixing condition $A_u=0$ in the flat space case, and which has a quite different structure.  We can find self-dual Yang-Mills fields in this gauge which satisfy a generalised Plebanski equation, for background metrics which are of course self-dual themselves, namely
\begin{equation}\label{eq:SDbackground}
{\rm Pleb}_0(\Phi)=0,
\end{equation}
but also are of the Kerr-Schild form and so satisfy
\begin{equation}\label{eq:KSPhi}
\Phi_{Xv}^2-\Phi_{XX}\Phi_{vv}=0.
\end{equation}
The above conditions imply that the flat Laplacian acting on $\Phi$ vanishes i.e.  $\Phi_{uv}=\Phi_{XY}$. The self-dual gauge field in this case is given by
\begin{equation}\label{eq:NewYM2}
\begin{split}
A_u & = \frac{1}{2}\Big(\phi_Y \Phi_{XX}\ - \phi_u\Phi_{Xv}\Big),\\
A_v &=\hat k_Y(\Phi)(\phi) := \phi_Y -\frac{1}{2}\Big(\phi_v \Phi_{Xv}\ - \phi_X\Phi_{vv}\Big)   ,\\
A_X  &= \hat k_u(\Phi)(\phi) :=\phi_u -\frac{1}{2}\Big(\phi_v \Phi_{XX}\ - \phi_X\Phi_{vX}\Big)  ,\\
A_Y &=  \frac{1}{2}\Big(\phi_Y \Phi_{Xv} - \phi_u\Phi_{vv}\Big) ,
\end{split}
\end{equation}
where the previously defined $\hat k_u$ and $\hat k_Y$  are now regarded as differential operators defined by replacing the unhatted $k$'s in their expression by the corresponding derivatives.
The perturbation field $\phi$ then satisfies
 a generalised Plebanski equation in the background $\Phi$ given by
\begin{equation}\label{eq:WaveGen}
\Box_\Phi \phi -2\Big[\hat k_u(\Phi)(\phi),\hat k_Y(\Phi)(\phi)\Big]=0 .
\end{equation}
The gauge field above is not adapted to the background metric in the same fashion as the previous gauge, instead it features the non-trivial components of the background metric. Since in the flat space limit it is related to the previous gauge by the coordinate exchange $u\leftrightarrow v$ and $Y\leftrightarrow X$ we call it the `\textit{flipped gauge}'.

The commutator term in \eqref{eq:WaveGen}
in this
`flipped gauge' reveals a different algebraic structure connected with the fact that one can define a curved space Poisson bracket for this spacetime  \cite{Gindikin:1986,Dunajski:2000iq}. 
We can define this by considering the expression $\hat k_{1u}\hat k_{2Y} - \hat k_{2u}\hat k_{1Y}$ and as before replacing the unhatted $k$'s in this expression by coordinate derivatives with respect to the two functions in the Poisson bracket, i.e.
\begin{equation}\label{eq:PBgen2}
\{f,g\}_{\Phi}=  \hat k_u(\Phi)(f)\hat k_Y(\Phi)(g)-\hat k_Y(\Phi)(f)\hat k_u(\Phi)(g)\,,
\end{equation}
or
\begin{equation}\label{eq:PBgen2Extra}
\begin{split}
\{ u,v\}_{\Phi} &= -\frac{1}{2}\Phi_{Xv}, \quad  
\{ u,X\}_{\Phi} = \frac{1}{2}\Phi_{vv}, \quad
\{ u,Y\}_{\Phi} =1,  \\
\{ v,X\}_{\Phi}& = \frac{1}{4}\Big( \Phi_{Xv}^2 - \Phi_{XX}\Phi_{vv}\Big), \quad
\{ v,Y\}_{\Phi} = -\frac{1}{2}\Phi_{XX}, \quad\{ X,Y\}_{\Phi} = \frac{1}{2}\Phi_{Xv} .
\end{split}
\end{equation}
The Jacobi identity for the Poisson bracket $\{ \, , \, \}_{\Phi}$ is satisfied since the self-dual background $\Phi$ satisfies the Plebanski equation ${\rm Pleb}_0(\Phi)=0$. Furthermore, since the Kerr-Schild condition $\Phi_{Xv}^2 - \Phi_{XX}\Phi_{vv}=0$ is satisfied the bracket $\{ v,X\}$ vanishes. The symplectic form connected with the Poisson bracket \eqref{eq:PBgen2} is
\begin{equation}\label{eq:symplectic}
\omega=\frac{1}{2}\Phi_{Xv}(du\wedge dv - dX\wedge dY)- \frac{1}{2}\Phi_{vv}\, dv\wedge dY + \frac{1}{2}\Phi_{XX} du\wedge dX + du\wedge dY \,.
\end{equation}
We observe that $\omega^2=0$ and is closed, $d\omega=0$ (c.f. \cite{Dunajski:2000iq}) when the Kerr-Schild condition and background Plebanski equation are satisfied.

Using the notation 
\begin{equation}\label{eq:Commutatorothergf}
[\{ f,g\}]_{\Phi} := \Big[ \hat k_u(\Phi)(f),\hat k_Y(\Phi)(g)\Big] - \Big[ \hat k_Y(\Phi)(f),\hat k_u(\Phi)(g)\Big] ,
\end{equation}
the condition \eqref{eq:WaveGen}  on the field $\phi$ may then be written 
\begin{equation}\label{eq:commutatorothergf}
\Box_{\Phi}\phi - [\{\phi,\phi\}]_{\Phi} =0 .
\end{equation}
%


\subsection{Self-dual gravity}\label{sec: gravityPertubation}

We now consider self-dual gravity perturbations on the background metric in \eqref{eq:GenMetric}. That is we simply consider the shifted metric $g_{\mu\nu}(\Phi+\Psi)$ given by \eqref{eq:GenMetric} with $\Phi$ replaced by $\Phi+\Psi$. We take the metric $g_{\mu\nu}(\Phi)$ to be the background self-dual spacetime, with ${\rm Pleb}_0(\Phi)=0$. This setup corresponds to the so called `matched gauge' for the gravity perturbation.

We can then define the gravitational Plebanski function in the background metric $g_{\mu\nu}(\Phi)$ by
\begin{equation}\label{eq:Plebeq}
{\rm Pleb}_{\Phi}(\Psi)\coloneqq \Box_{\Phi}\Psi+ \Psi_{Xv}^2 - \Psi_{XX}\Psi_{vv} ,  
\end{equation}
with $\Box_\Phi$ the scalar Laplacian in the background metric. Once again this can be written in terms of the double bracket notation \eqref{eq:PoissonBgrav}
\begin{equation}\label{eq:Plebeqgravbrac}
{\rm Pleb}_{\Phi}(\Psi) = \Box_{\Phi}\Psi-\{\{\Psi, \Psi\}\}\, , 
\end{equation}
illustrating the double copy structure compared with eqn. \eqref{eq:PlebYMEHPB}. Alternatively, \eqref{eq:Plebeqgravbrac} can be viewed as the double copy of \eqref{eq: YMplebanskiYMBackground}, where we double copy both the YM background and the perturbation.
Now one can check that the Plebanski equation satisfies the following identity
\begin{equation}\label{eq:PlebID}
{\rm Pleb}_{0}(\Phi+\Psi) =  {\rm Pleb}_{0}(\Phi) + {\rm Pleb}_{\Phi}(\Psi)  .
\end{equation}
This immediately gives the gravitational Plebanski equation in the background metric as simply
\begin{equation}\label{eq:Plebeq2}
{\rm Pleb}_{\Phi}(\Psi) = 0 .  
\end{equation}
This follows since the identity \eqref{eq:PlebID} shows that  if $\Phi$ leads to a self-dual metric then $\Phi+\Psi$ does as well if the Plebanski equation for $\Psi$ in a $\Phi$ metric background  \eqref{eq:Plebeq2}  is satisfied. The above conclusions can be confirmed explicitly. The relevant non-trivial component of the anti-self-dual  part of the Weyl tensor for the metric $g_{\mu\nu}(\Phi+\Psi)$ is given by
\begin{equation}\label{eq:GravASDWeylDouble}
\begin{split}
C^-_{uYuY}(g_{\mu\nu}(\Phi+\Psi)) &=  -\frac{1}{4}\Delta_0(\Phi+\Psi) {\rm Pleb}_0(\Phi+\Psi)) \\
    &=   -\frac{1}{4}\Delta_0(\Phi+\Psi) (  {\rm Pleb}_{0}(\Phi) + {\rm Pleb}_{\Phi}(\Psi)  ) \\
     &=   -\frac{1}{4}\Delta_0(\Phi+\Psi) ( {\rm Pleb}_{\Phi}(\Psi)  )\\
     & =0 
\end{split}
\end{equation}
where we have used the self-duality of the background metric, with $  {\rm Pleb}_{0}(\Phi) =0$, and imposed the condition \eqref{eq:Plebeq2}. 
It is also immediate that the variations of the Plebanski function \eqref{eq:Plebeq} are related to variations of the flat Plebanski function -  if we define the variation
\begin{equation}
    \Delta_\Phi(\Psi)(\delta\Psi) := \delta_{\Psi}({\rm Pleb}_\Phi(\Psi)) \,,
\end{equation}
then as differential operators
\begin{equation}
    \Delta_\Phi(\Psi)=\Delta_0(\Phi+\Psi)\,,
\end{equation}
as expected. A similar argument, based on \eqref{eq:RicciFlat}, shows Ricci-flatness of the shifted metric.

In summary, we have shown the following commuting triangle of double copy relations for equations of motion in the matched gauge:
\begin{equation}\label{fig: double copies}
    \begin{tikzcd}[column sep=large,row sep=large]
    {\text{SDYM on SDYM }}\atop{\text{\eqref{eq: deformedLaplacian}}} \arrow{r}  {{\text{Double copy}}\atop{\text{Background}}} \arrow{dr}
    &  {\text{SDYM on SDG}}\atop{\text{ \eqref{eq:PlebYMEHPB}}} \arrow{d}{{\text{Double copy}}\atop{\text{Perturbation}}}\\
    & {\text{SDG on SDG }}\atop{\text{ \eqref{eq:Plebeq2}}}
    \end{tikzcd}\,.
\end{equation}
The diagonal arrow above is just the usual self-dual flat space double copy applied to the sum of the background and perturbation fields $\chi+\psi$ in \eqref{eq: YMbackYMpertubation}. The double copy properties of backgrounds and perturbations have been studied beyond the self dual context in \cite{Bahjat-Abbas:2017htu}.

We can also consider the `flipped gauge' for which the natural double copy of the bracket in \eqref{eq:Commutatorothergf} replaces the YM commutator with the  Poisson brackets $\{\, ,\,\}_\Phi$ of \eqref{eq:PBgen2}\footnote{Other double brackets may be defined by dropping the $\Phi$ terms inside the brackets in \eqref{eq:Doublebkt2} and/or using the flat space Poisson bracket.}
\begin{equation}\label{eq:Doublebkt2}
\{\{ f,g\}\}_{\Phi} := \frac{1}{2}\bigg(
 \Big\{ \hat k_u(\Phi)(f),\hat k_Y(\Phi)(g)\Big\}_{\Phi} - \Big\{ \hat k_Y(\Phi)(f),\hat k_u(\Phi)(g)\Big\}_{\Phi} \bigg)\,.
\end{equation}
These double brackets have a related curved space Plebanski equation of the  form 
\begin{equation}\label{eq:DoublebktEqnzz}
\Box_\Phi\Psi -  
\{\{\Psi  , \, \Psi\}\}_\Phi = 0 ,
\end{equation}
which may be regarded as the double copy of \eqref{eq:commutatorothergf}. We discuss these brackets further in the examples below. It would be interesting to know if  these equations are related to the conditions required for the self-duality of the curvature of metrics on self-dual backgrounds.
One might also  study  self-dual backgrounds satisfying the Kerr-Schild condition $\Phi_{Xv}^2=\Phi_{XX}\Phi_{vv}$ more generally. Whilst we have not found answers to these questions in the general case, the study of interesting examples reveals more structure, as we will see in the following.


\section{The self-dual plane wave spacetime}\label{sec: SDPWBackground}

Plane wave backgrounds have been the object of some interest recently in the area of amplitudes, kinematic algebras and the double copy (see, for example, \cite{Adamo:2017nia,
 Adamo:2018mpq, Adamo:2020yzi, Adamo:2022mev, Adamo:2019zmk, Adamo:2017sze, Adamo:2021hno}
and references therein).
Here we study the self-dual plane wave metric
\begin{equation}\label{eq:PWmetric}
ds^2_{PW} = 2 du dv - 2 dX dY + 2 F(v)dY^2 ,
\end{equation}
where $F(v)$ is a function related to the wave profile. 

This metric is an example of the general form \eqref{eq:GenMetric} considered earlier and is also Kerr-Schild, we simply set $\Phi=\Phi(v)$ with $\Phi_{vv}=2 F(v)$.
The self-dual plane wave metric is Ricci-flat and has self-dual Weyl tensor;  the only non-vanishing components of the self-dual part of the Weyl tensor being $C_{vYvY} = -2 F''[v]$ and those related to this by the symmetries of this tensor.

\subsection{Self-dual Yang-Mills}

A self-dual gauge field in the `matched gauge' on this spacetime is given by
\begin{equation}\label{eq:PWsdym}
A_{\mu}=(A_u,A_v,A_X,A_Y) = (\phi_X,0,0,\phi_v) ,
\end{equation}
where, in order to solve the self-duality conditions, the scalar field $\phi(u,v,X,Y)$ must satisfy the plane wave background Plebanski equation
\begin{equation}\label{eq:PWymeqn}
\Box_{PW}\phi - 2[\phi_v,\phi_X]= 0 ,
\end{equation}
with $\Box_{PW}$ the Laplacian in the metric \eqref{eq:PWmetric}.
Using the Poisson bracket $\{ f, g\} = f_v g_X - f_X g_v$ which is the same as the flat space case, we can write \eqref{eq:PWymeqn} as
\begin{equation}\label{eq:PWymeqn2}
\Box_{PW}\phi - [\{\phi,\phi\}]= 0 ,
\end{equation}
where the double bracket notation \eqref{compoisbracket} is defined as usual.

There is also the `flipped' self-dual gauge field solution in this background, from \eqref{eq:NewYM2} which can be used to elucidate the algebraic structure of self-dual perturbations on the self-dual plane wave background. We find
\begin{equation}\label{eq:PWsdymflipped}
A_{\mu}=\big(0,\phi_Y + F(v)\phi_X,  \phi_u, -F(v)\phi_u  \big) ,
\end{equation}
where the field $\phi$  satisfies 
\begin{equation}\label{eq:PWymeqnflipped}
\Box_{PW}\phi - 2[\phi_u,\phi_Y+F(v)\phi_X]= 0 .
\end{equation}
This leads us to the modified Poisson bracket in the plane wave background
\begin{equation}\label{eq:PWpbflipped}
\{ f, g\}_{PW} = f_u \big(g_Y+F(v)g_X\big) - \big(f_Y+F(v)f_X\big) g_u ,
\end{equation}
and the re-writing of \eqref{eq:PWymeqnflipped} as 
\begin{equation}\label{eq:PWymeqnflippedDC}
\Box_{PW}\phi - [\{\phi,\phi\}]_{PW}= 0 .
\end{equation}

We are now tasked with finding the analogue of plane wave solutions to the wave equation in flat space, but for solutions to the wave equation in the background \eqref{eq:PWmetric}. Such solutions then act as generators of our kinematic Poisson algebra. We begin by constructing a null vector in flat space $k_{\mu}$ satisfying $k_u k_v-k_Xk_Y=0$ so that (as before) $k_u=\rho k_X, k_Y=\rho k_v$ for some $\rho$. Then for the function $G(v)$ given by the indefinite integral of $F(v)$, ie $G'=F$, we may define the quantity
\begin{equation}\label{eq:PWepguy}
\begin{split}
Q_k (u,v,X,Y) &\coloneqq (\rho Y +  v)k_v + (\rho u+ X)k_X + \frac{1}{\rho} G(v) k_X\\
&=k \cdot x + \frac{1}{\rho} G(v) k_X\, . 
\end{split}
\end{equation}

Then one can show that the vector $K_\mu=\nabla_\mu Q_k$ is null, $K^\mu K_\mu=0$, divergence free, $\nabla^\mu K_\mu=0$ (which is just the wave equation on $Q_k$), and geodesic, $K^\nu\nabla_\nu K_\mu=0$, where $\nabla_\mu$ is the covariant derivative in the plane wave metric. One consequence is that any function of $Q_k$ is  annihilated by the Laplacian, in particular
\begin{equation}\label{eq:PWepguy2}
\Box_{PW} e^{i Q_k (u,v,X,Y)}= 0 . 
\end{equation}
Whence the function $e^{i Q_k (u,v,X,Y)}$ satisfies the wave equation in the plane wave background and furthermore reduces to the usual plane wave $e^{i k \cdot x}$ in the flat space limit.
The Poisson bracket of two of these solutions is

\begin{equation}\label{eq:PWpb2}
\begin{split}
\{ e^{iQ_1},e^{iQ_2} \} &= e^{i(Q_1+Q_2)} (k_{1X} k_{2v}-k_{1v}k_{2X} +k_{1X}k_{2X}\frac{(\rho_1-\rho_2)}{\rho_1\rho_2}F(v))\\
&=:e^{i(Q_1+Q_2)} X_{PW}(k_1, k_2),
\end{split}
\end{equation}
leading to a modification of the structure constants defining the kinematic algebra compared to the flat space case. This modification is however sub-leading in the holomorphic collinear limit so we expect it to not alter the $w$-algebra, which we confirm in the next section. 

This result, and hence also the $w$-algebra in \eqref{eq:PWwalg}, also holds if one uses the flipped gauge Poisson bracket \eqref{eq:PWpbflipped},  although in that case it is more natural to write the function \eqref{eq:PWepguy} in terms of $k_u$ and $k_Y$ as follows
\begin{equation}
   Q_k (u,v,X,Y) = (Y + \tilde{\rho} v)k_Y + ( u+\tilde{\rho} X)k_u + \tilde{\rho}^2 G(v) k_u
\end{equation}
where $\tilde{\rho}\coloneqq 1/\rho$. The flipped Poisson bracket of two plane waves is then 
\begin{equation}
\begin{split}\label{eq:PWpb2Flipped}
    \{ e^{iQ_1},e^{iQ_2} \}_{PW} &= e^{i(Q_1+Q_2)} (k_{1Y}k_{2u}-k_{1u}k_{2Y} +k_{1u}k_{2u}(\tilde{\rho}_1-\tilde{\rho}_2)F(v)) \\
    &= e^{i(Q_1+Q_2)} (-\rho_1 \rho_2) X_{PW}(k_1, k_2)\,.
\end{split}
\end{equation}
\subsection{Self-dual gravity}
 
The gravitational analogue of the discussion above in the `matched gauge' is based on the metric
\begin{equation}\label{eq:PWgravmetric}
ds^2_{PWG} = 2 du dv - 2 dX dY + 2 F(v)dY^2 + \Psi_{vv}dY^2 + \Psi_{XX}du^2 + 2\Psi_{vX}du dY.
\end{equation}
This metric has vanishing Ricci tensor and self-dual Weyl tensor if the following Plebanski equation is satisfied:
\begin{equation}\label{eq:PWpleb}
\Box_{PW}\Psi +\Psi_{vX}^2-\Psi_{vv}\Psi_{XX} = 0 ,
\end{equation}
where $\Box_{PW}$ is the Laplacian in the self-dual background.
Employing the double bracket notation \eqref{eq:PoissonBgrav} as usual we have
\begin{equation}\label{eq:PWpleb2}
\Box_{PW}\Psi - \{\{\Psi,\Psi\}\} = 0 ,
\end{equation}
revealing the double copy structure compared with \eqref{eq:PWymeqn}.

If we apply the double bracket to two of the `plane waves' \eqref{eq:PWepguy} we find
\begin{equation}\label{eq: PWMatchedDoubleBracket}
    \{\{ e^{iQ_1},e^{iQ_2} \}\}=e^{i(Q_1+Q_2)} \frac{1}{2}\bigg(X_{PW}(k_1,k_2)^2-\frac{i k_{1X}k_{2X}(k_{1X}\rho_1+k_{2X}\rho_2)F'(v)}{\rho_1 \rho_2}\bigg)
\end{equation}
which is not just the simple square of the relation \eqref{eq:PWpb2}. Despite this, we can still derive a $w-$algebra as follows.  Similarly to the flat space case,  we may expand the above solutions to the wave equation in powers of soft momenta variables $k_v, k_X$ to find 
\begin{equation}\label{eq:PWwgens}
e^{i Q_k (u,v,X,Y)} = \sum_{a,b=0}^{\infty} \frac{(ik_v)^a(ik_X)^b}{a!b!} e_{ab}\,,
\end{equation}
where we have defined $e_{ab}= (\rho Y+ v) ^a (\rho u +  X+\frac{1}{\rho} G(v))^b$ in the self-dual plane wave background. Defining the modified $w$ generators
\begin{equation}\label{eq: wgenssSDPW}
w^p_m\coloneqq\frac{1}{2}e_{p-1+m,p-1-m}= \frac{1}{2}(\rho Y+ v) ^{p-1+m} (\rho u +  X+\frac{1}{\rho} G(v))^{p-1-m}\,,
\end{equation}
in analogy with the flat space case. We recover the standard $w_{1+\infty}$-algebra for these modified generators, working to leading order in the holomorphic collinear limit
\begin{equation}\label{eq:PWwalg}
\{w_m^p,w_n^q\}=  \Big( m(q-1) - n(p-1)\Big)w_{m+n}^{p+q-2}\,.  
\end{equation}

We may also consider the `flipped gauge' with its modified Poisson bracket \eqref{eq:PWpbflipped} which satisfies a double copy relation analogous to \eqref{eq:Gravbracketeikx} acting on two solutions $e^{i Q_k (u,v,X,Y)}$. First we define a modified double bracket
\begin{equation}\label{eq:PWdoublebkt}
\{\{f,g\}\}_{PW}= \frac{1}{2}\Big(\{f_u , g_Y+F(v)g_X\}_{PW} - \{ f_Y+F(v)f_X, g_u\}_{PW} \Big) \, ,
\end{equation}
then we find the expected double copy of \eqref{eq:PWpb2Flipped}, that is
\begin{equation}\label{eq:PWpbPW2}
\{\{ e^{iQ_1},e^{iQ_2} \}\}_{PW} = \frac{1}{2}e^{i(Q_1+Q_2)} (-\rho_1\rho_2 X_{PW}(k_1, k_2))^2 . 
\end{equation}
Interestingly, in contrast to the matched double bracket \eqref{eq: PWMatchedDoubleBracket} of plane waves, the above does exhibit a simple squaring relation when compared to the single bracket  \eqref{eq:PWpb2Flipped}. As mentioned before, we can also define analogous soft generators $\tilde{w}_m^p$ in the flipped gauge, now as coefficients of $k_u^a$ and $k_Y^b$. One can then show that these generators also satisfy the $w_{1+\infty}$ algebra \eqref{eq:PWwalg}, but now with the bracket \eqref{eq:PWpbflipped}.


\section{The Eguchi-Hanson spacetime}\label{sec: EHBackground}
We now move on to consider a more complicated example, the Eguchi-Hanson space-time. This is self-dual, and in the form \eqref{eq:SDbackground} has the scalar function
\begin{equation}\label{eq:EHphi}
\Phi_{EH} = \frac{m v^2}{2Y^2(u v - X Y)} , 
\end{equation}
with $m$ a constant, satisfying the Plebanski equation in flat space
\begin{equation}\label{eq:PlebEHphi}
{\rm Pleb}_0(\Phi_{EH}) = 0.
\end{equation}
The full metric is then
\begin{equation}\label{eq:EHmetric}
\begin{split}
ds^2_{EH} & = g_{\mu\nu}(\Phi_{EH})dx^\mu dx^\nu \\
&= 2 du dv - 2 dX dY + \frac{m v^2}{(uv-XY)^3}du^2 +\frac{m X^2}{(uv-XY)^3}dY^2 -\frac{2m vX}{(uv-XY)^3}du dY 
\end{split}
\end{equation}
and satisfies the Kerr-Schild condition.

We now repeat the methods laid out for the general case and the plane wave example but now with the function $\Phi_{EH}$. We  will encounter a much richer algebraic structure than was found in the self-dual plane wave background, reproducing in spacetime some of the results recently described via twistor space in \cite{Bittleston:2023bzp}.

\subsection{Self-dual Yang-Mills}

Consider firstly self-dual Yang-Mills in an Eguchi-Hanson background. From the results earlier, a gauge field $A_\mu$ in the `matched gauge' $A_v=0$ has self-dual field strength if in addition $A_X=0, A_u =  \phi_X$ and
$A_Y =  \phi_v$, with $\phi$ satisfying the Plebanski equation in the EH background
\begin{equation}\label{eq:PlebYMEH}
\Box_{EH}\phi -2[\phi_v,\phi_X] = 0 ,
\end{equation}
where $\Box_{EH}$ is the Laplacian in the metric \eqref{eq:EHmetric}. In the case at hand, the EH Plebanski equation can be written in terms of the flat space Poisson bracket \eqref{eq:PoissonBYMflat} as
\begin{equation}\label{eq:PlebYMEHpoisb}
\Box_{EH}\phi -[\{\phi,\phi\}] = 0 \,.
\end{equation}

The alternative `flipped gauge' \eqref{eq:NewYM2} in the Eguchi-Hanson case comes from the null vector $m^\mu = (1,0,0,v/X)$ and gauge-fixing condition $m^\mu A_\mu=0$ and sets
\begin{equation}\label{eq:AvXY}
\begin{split}
A_u&=\frac{m v}{2(u v - X Y)^3}\Big( v\phi_Y + X \phi_u \Big)  ,\\
A_v & = \hat k_Y(\Phi_{EH})(\phi)=\phi_Y + \frac{m X}{2(u v - X Y)^3}\Big( X\phi_X + v \phi_v \Big) , \\
A_X & = \hat k_u(\Phi_{EH})(\phi)= \phi_u - \frac{m v}{2(u v - X Y)^3}\Big( X\phi_X + v \phi_v \Big) , \\
A_Y & =- \frac{m X}{2(u v - X Y)^3}\Big( v\phi_Y + X \phi_u \Big) .
\end{split}
\end{equation}
This gauge field has self-dual field strength if the scalar field $\phi$ satisfies
\begin{equation}\label{eq:EHothergf}
\Box_{EH}\phi - 2\Big[\hat k_u(\Phi_{EH})(\phi)\, ,\, \hat k_Y(\Phi_{EH})(\phi)\Big]=0 .
\end{equation}
Using the notation 
\begin{equation}\label{eq:CommutatorothergfEH}
[\{ f,g\}]_{EH} := \Big[f_u-\frac{m v}{2(u v - X Y)^3}\big(X f_X+v f_v\big)\,  , \, g_Y+\frac{m X}{2(u v - X Y)^3}\big(X g_X+v g_v\big)\Big] + (f\leftrightarrow g) ,
\end{equation}
equation \eqref{eq:EHothergf} may be written (c.f. \eqref{eq:PlebYMEHPB})
\begin{equation}\label{eq:EHcommutatorothergf2}
\Box_{EH}\phi - [\{\phi,\phi\}]_{EH} =0 .
\end{equation}
The deformed Poisson bracket \eqref{eq:PBgen2} in the Eguchi-Hanson metric is then
\begin{equation}\label{eq:PBothergf}
\begin{split}
\{f,g\}_{EH} &= \hat k_u(\Phi_{EH})(f) \,\hat k_Y(\Phi_{EH})(g)- \hat k_Y(\Phi_{EH})(f) \,\hat k_u(\Phi_{EH})(g)\,, \\
\end{split}
\end{equation}
and we note that the terms quadratic in $m$ in the above Poisson bracket in fact drop out.

To find the equivalent of plane wave solutions in the EH background we introduce a null co-vector $k_\mu=(k_u,k_v,k_X,k_Y)$ whose components satisfy $k_uk_v=k_Xk_Y$ so that as before we may write 
\begin{equation}\label{eq:rhos}
\frac{ k_u}{k_X} = \frac{k_Y}{k_v} = \rho
\end{equation}
for some parameter $\rho$. As was the case for the self-dual plane wave, we look for solutions to the EH wave equation which are of an exponential form and return the usual $e^{ik \cdot x}$ plane wave in the flat space limit $m\rightarrow 0$.
Following \cite{Bittleston:2023bzp}, we define the function 
\begin{equation}\label{eq:curious}
\begin{split}
R_k (u,v,X,Y) &\coloneqq (k\cdot x)^2 - \frac{m (v k_v + X k_X)^2}{2(uv-XY)^2} \\
 & = ((\rho Y+ v)k_v+(\rho u+ X)k_X)^2 - \frac{m (v k_v + X k_X)^2}{2(uv-XY)^2}.
\end{split}
\end{equation}
Then the vector $K_\mu=\nabla_\mu R_k$ is null, $K^\mu K_\mu=0$, divergence free, $\nabla^\mu K_\mu=0$ (which is just the wave equation on $R_k$), and geodesic, $K^\nu\nabla_\nu K_\mu=0$, where $\nabla_\mu$ is the covariant derivative in the EH metric.
One consequence is that any function of $R_k$ is  annihilated by the Laplacian, in particular  
\begin{equation}\label{eq:curiouser}
\Box_\Phi e^{i\sqrt{R_k (u,v,X,Y)}} = 0 ,
\end{equation}
where $e^{i\sqrt{R_k (u,v,X,Y)}}$ gives the standard plane wave $e^{i k\cdot x}$ in the flat space limit $m\rightarrow 0$. Note the qualitative difference between the Eguchi-Hanson function $R_k$, which is quadratic in the null momenta $k_{\mu}$, versus $Q_k$ in the self-dual plane wave background which is linear in $k_{\mu}$. 

 We can now perform the Poisson bracket of two of the solutions $e^{i\sqrt{R_k}}$ to the wave equation with momenta $k_1,k_2$, using the form of $R_k$ on the second line of \eqref{eq:curious}. We work in the holomorphic collinear limit $\rho_1=\rho_2=\rho$ which is all that is needed to recover a $w$-algebra. This gives
\begin{equation}\label{eq:Expothergf}
\{ e^{i\sqrt{R_1}}, e^{i\sqrt{R_2}} \} = 
e^{i(\sqrt{R_1}+\sqrt{R_2})} X_{EH}(k_1,k_2) 
\end{equation}
where here
\begin{equation}\label{eq:DCmaybe}
X_{EH}(k_1,k_2)= \frac{1}{\sqrt{R_1R_2}}\Big(k_{1X}k_{2v}-k_{1v}k_{2X}\Big)\Big((k_1\cdot x)(k_2\cdot x) - \frac{m(v k_{1v} + X k_{1X})(v k_{2v} + X k_{2X})}{2(uv-XY)^2}\Big) ,
\end{equation}
and dot products $k\cdot x$ here mean $(\rho Y+ v)k_v+(\rho u+ X)k_X$. The final factor may be compared to the right-hand side of \eqref{eq:curious}. Eqn \eqref{eq:Expothergf} may be viewed as the Eguchi-Hanson background version of the expression in equation \eqref{eq:cubicX2}. We note that the kinematic algebra has modified kinematic structure `constants' compared to the flat-space case and the modification survives in the holomorphic collinear limit so we expect the $w$-algebra to also be modified. As in the plane wave case, the  Poisson bracket relation \eqref{eq:Expothergf} also holds if we use the Eguchi-Hanson flipped bracket \eqref{eq:PBothergf}, up to an overall factor which also appeared in  \eqref{eq:PWpb2Flipped}.

\subsection{Self-dual gravity}

For the case of self-dual gravity, a perturbation of the EH metric in the matched gauge is given by $g_{\mu\nu}(\Phi_{EH}+\Psi)= g_{\mu\nu}(\Phi_{EH})+g_{\mu\nu}(\Psi)$ has vanishing anti-self-dual components of the Weyl tensor except for
\begin{equation}\label{eq:GravASDWeyl2}
C^{EH-}_{uYuY}= -\frac{1}{4}\Delta_{\Phi_{EH}}(\Psi) ( {\rm Pleb}_{\Phi_{EH}}(\Psi) ) .
\end{equation}
Thus, the perturbed EH metric  has self-dual Weyl tensor if the EH Plebanski equation is satisfied. The EH Plebanski equation for self-dual gravity in this case is given by
\begin{equation}\label{eq:PlebGrav2}
{\rm Pleb_{\Phi_{EH}}}(\Psi)= \Box_{EH}\Psi -\Psi_{XX}\Psi_{vv} + (\Psi_{vX})^2=\Box_{EH}\Psi-\{\{\Psi, \Psi\}\}= 0 \,,
\end{equation}
using the double bracket \eqref{eq:PoissonBgrav}.
Similarly for  the Ricci tensor one finds that its components vanish except for $R^{EH}_{ab}$ with $a, b\in (u,Y)$ and for these components
\begin{equation}\label{eq:Ricci}
R^{EH}_{ab}= - \frac{1}{2}\partial_{\bar a}\partial_{\bar b} {\rm Pleb}_{\Phi_{EH}}(\Psi) ,
\end{equation}
where $\bar u = X, \bar Y = v$. 

The non-trivial form of the single bracket \eqref{eq:Expothergf} suggests that the double copy, realised by using a double bracket, may involve more than just the square of $X(k_1,k_2)$. This proves to be the case - the double brackets of two plane wave solutions $e^{i\sqrt{R_k}}$ in the EH background in the holomorphic collinear limit give a double copy-type formula
\begin{equation}\label{eq:doublebktwaves}
\{\{ e^{i\sqrt{R_1}}, e^{i\sqrt{R_2}} \}\}= \frac{1}{2} e^{i(\sqrt{R_1}+\sqrt{R_2})}X_{EH}(k_1,k_2)^2 + \dots ,
\end{equation}
(c.f \eqref{eq:Gravbracketeikx} in the flat space case) where $X_{EH}(k_1,k_2)$ is given in \eqref{eq:DCmaybe} and the terms indicated by dots are more complicated expressions which multiply $(R_1)^{-1/2}, (R_2)^{-1/2}$ and $(R_1R_2)^{-1/2}$ and are of order $m, m^2$ or $m^3$ and hence vanish in the flat space limit $m\rightarrow 0$. These results suggest that in general the double copy and related kinematic algebra on curved space backgrounds are not just given by a simple squaring operation of the relevant curved space term, as seen in the first term on the right-hand side of eqn. \eqref{eq:doublebktwaves}, but can involve other curvature corrections.

We now consider the soft expansion of the solution $e^{i\sqrt{R_k (u,v,X,Y)}}$
 in powers of the soft momentum variables $k_Y,k_u$ and once again work in the holomorphic collinear limit where $\rho_1 =\rho_2=\rho$. We define functions $X_g, Y_g, Z_g$ which give the coefficients of $k_v^2, k_X^2$ and $k_v k_X$ in the function $R_k$
\begin{equation}\label{eq:XYZ}
\begin{split}
X_g &= (\rho Y +  v)^2 - \frac{m  v^2}{2(u v - X Y)^2} \,, \\
Y_g &= (\rho u +  X)^2 - \frac{mX^2}{2(u v - X Y)^2}\, , \\
Z_g &= ( \rho Y +  v)( \rho u +  X) - \frac{m X v}{2(u v - X Y)^2}\, , \\
\end{split}
\end{equation}
which satisfy
\begin{equation}\label{eq:curious2}
X_g k_v^2 + Y_g k_X^2 + 2 Z_g k_v k_X = (k\cdot x)^2 - \frac{m(v k_v + X k_X)^2}{2(uv-XY)^2}=R_k\,,
\end{equation}
and the discriminant constraint
\begin{equation}\label{eq:XYZid}
X_gY_g - Z_g^2 = -\frac{m\rho^2}{2}\, .
\end{equation}
The quantities $X_g,Y_g,Z_g$ correspond to the $X,Y,Z$ of \cite{Bittleston:2023bzp}. The parameter $c^2(\lambda)$ in that reference is related to ours by $c^2(\lambda)=\frac{m^2\rho^2}{2}$.

One can then expand the \lq plane wave\rq\ $e^{i\sqrt{R_k}}$ in powers of the variables $k_v,k_X$\footnote{The authors of \cite{Bittleston:2023bzp} consider the quantity $\cos(\sqrt{R_k})$ since they also impose the $Z_2$ symmetry required by global considerations. This involves the same basic generators.} and the  Poisson bracket of the coefficients in this expansion generates a $w$-type algebra. 
Due to the  constraint \eqref{eq:XYZid} one can define a new basis of generators 
$V_{2p,2q}:=X_g^pY_g^q,\,  V_{2p+1,2q+1}:=X_g^pY_g^qZ_g $, and the Poisson brackets of these generates the underlying algebra 
\begin{equation}\label{eq:XYZPB2}
\begin{split}
\{ V_{2p,2q},V_{2r,2s} \} &= 4(ps-qr)V_{2p+2r-1,2q+2s-1} \,,\\
\{ V_{2p,2q},V_{2r+1,2s+1}\} &= 2(p(2s+1)-q(2r+1))V_{2p+2r,2q+2s} 
+ 2 m \rho^2 (ps-qr)V_{2p+2r-2,2q+2s-2}\, , \\
\{ V_{2p+1,2q+1},V_{2r+1,2s+1}\} &= ((2p+1)(2s+1)-(2q+1)(2r+1))V_{2p+2r+1,2q+2s+1 }    \\
&\qquad \qquad \qquad + 2 m \rho^2
 (ps-qr)V_{2p+2r-1,2q+2s-1} \,.
\end{split}
\end{equation}
The full celestial chiral algebra of self-dual gravity on an Eguchi-Hanson background can then be written in terms of sums of these generators (see \cite{Bittleston:2023bzp}).

We can also consider the double brackets of the flipped gauge \eqref{eq:Doublebkt2} in the EH background which are given by 
\begin{equation}\label{eq:Doublebkt2EH}
\{\{ f,g\}\}_{EH} := \frac{1}{2} 
\Big\{f_u-\frac{m v}{2(u v -X Y)^3}\big(v f_v\ + X f_X\big)\,  , \, g_Y+\frac{m X}{2(uv-XY)^3}\big(v g_v+ Xg_X\big) \Big\}_{EH} + (f\leftrightarrow g) \,,
\end{equation}
and using these in the holomorphic collinear limit we find a double copy-type formula like \eqref{eq:doublebktwaves} with the same leading term, but with different sub-leading terms. As in the plane wave case, we could expand the solution \eqref{eq:curious} in terms of $k_u^2$, $k_Y^2$ and $k_{u}k_Y$ instead to define analogous soft generators $\tilde{X}_g$, $\tilde{Y}_g$ and $\tilde{Z}_g$. These then satisfy the same algebra as \eqref{eq:XYZPB2} but with the flipped Poisson bracket \eqref{eq:PBothergf}.
%


\section{Conclusions}

We have studied the self-duality of gauge and gravitational fields on the self-dual background spacetimes defined by solutions of Plebanski's second equation. In light-cone gauges we showed that the conditions for self-duality could be reduced to second order scalar equations generalising the flat space equations. We found two classes of general solutions.
One, which we called a \lq matched\rq\ gauge, was a direct generalisation of the flat space solutions to the curved self-dual backgrounds under consideration. The other involves a Kerr-Schild condition on the gravitational background, which we called  the \lq flipped\rq\ gauge, and can be seen as the curved space versions of \lq flipped\rq\ flat space solutions.  We discussed the double copy and kinematic algebra in these two cases. Finally, we studied  two examples in more detail - the self-dual plane wave spacetime
and the Eguchi-Hanson (EH) metric -  connecting with  some recent results from \cite{Adamo:2022mev} and \cite{Bittleston:2023bzp}, and noting that in the  EH background the kinematic algebra squaring relations are modified by curvature terms.

There are a number of avenues of research which  follow from this. It would be interesting to explore more examples in detail, and investigate perturbative solutions to the equations
where direct solutions prove difficult. Gravitational analogues of the \lq flipped\rq\,
gauge self-dual YM solution, eqn. \eqref{eq:NewYM2} could be studied further, in general and in particular examples. Plebanski-type conditions of the generic form $\Box\phi-\{\{\phi,\phi\}\}=0$ for the different double brackets given above would be expected to feature. In radiative spacetimes this should connect with the very recent analysis of self-dual deformations  in \cite{Adamo:2022mev}, which relates these to twistor sigma models and MHV generating functionals.
 It would also be interesting to explore applications to known deformations of the Plebanski equations such as those involving Moyal brackets (c.f.
\cite{Chacon:2020fmr, Bittleston:2023bzp} and references therein). The application of the formalism used recently for self-dual YM in \cite{Bonezzi:2023pox} could also be explored in self-dual backgrounds.

\vspace{24pt}

{\bf Acknowledgements:} We would like to thank 
Ricardo Monteiro, Chris White and Sam Wikeley for helpful comments.
This work was supported by the Science and Technology Facilities Council (STFC) Consolidated Grants ST/P000754/1 “String theory, gauge theory and duality” and ST/T000686/1 “Amplitudes, strings and duality”. The work of GRB and JG is supported by STFC quota studentships.


\bibliographystyle{utphys}
\bibliography{main}
\end{document}